# Dynamic Electrophysical Characterization of Porous Silicon based Humidity Sensitive Elements

S. Bravina, N. Morozovsky, R. Boukroub[*]

Institute of Physics of NAS of Ukraine, 46 Prospect Nauki, 03028 Kiev, Ukraine
*e-mail: bravina@iop.kiev.ua*
[*]Interdisciplinary Research Institute, IRI, IEMN-IRI, Avenue Poincaré - BP 69
59652 Villeneuve d'Ascq, France

The results of the investigation of changes of parameters of dynamic bipolar charge-voltage and bipolar and unipolar dynamic current-voltage characteristics and transient currents connected with the pulse change of humidity for the samples of por-Si are presented.

The view of high voltage current-voltage curves is characteristic for poling processes in the space charge region similar to that observed in the case of typical ionic conductors.

Observed phases of transformation of investigated electrophysical characteristics reflect the time scale of processes in the consequence "adsorption-dissociation and transfer – desorption".

The efficiency of using the methods of dynamic electrophysical characterization for studying characteristics of porous materials under fast humidity changes was demonstrated.

The problem of environment parameters monitoring demands from the modern sensorics many efforts directed to development of sensitive, high speed and stable humidity sensors integrated in the modern silicon (Si) base.

Electrophysical and thermal characteristics of systems based on porous materials with open pores are rather sensitive to the nature of molecules penetrated into the pores [1]. That is the base of using porous materials in sensors of environmental control, in particular for humidity sensors.

In this way many humidity sensitive media with open pores were examined. As promising materials are selected the following: polymers as the cheapest, porous ceramics as more flexible, and zeolite-like systems and also mezo-porous phases as thermo-stable.

Recently we have investigated the humidity-electric activity in zeolite-like systems and mezo-porous phases [2] and some of porous ceramics [3].

Taking into account a stable tendency of integrating environmental sensors into Si-based modern microelectronics it is reasonable to perform the investigations of humidity-electric activity in porous silicon (PSi) [4]. This material is considered as a promising one due to its photo- and electro - luminescence properties as well as due to its thermal properties, namely our thermowave-probing investigations confirmed the prospects of using of PSi in thermal sensorics as a buffer insulating substrate material of integrated thermal detectors [5]. We also marked in [5] that the application of por-Si is also possible for design of elements of heat-insulating surrounding of temperature-controlled units of electronic devices based on Si-technology.

On account of the well known humidity sensing of PSi [6] and the criticality of humidity level on electronic components operation the investigations of humidity impact on electrophysical characteristics of PSi in pulse mode under environmental conditions similar to ones of operating Si-based devices are desirable.

In this paper we present the results of investigating the changes of parameters of bipolar and unipolar dynamic current-voltage characteristics and transient currents connected with the pulse change of humidity for the samples of porous Si.

## 2. EXPERIMENT

### 2.1. Samples

The samples of PSi were prepared on 500 μm of thickness Si(111) B-doped p-type wafers with resistivity $\rho$ = 6-10 Ohm·cm. Anodic electrochemical etching was performed in electrolyte HF/EtOH=1/1 with current density of anodization 5-20 mA/cm$^2$ during 5-8 min. Then the samples were aged in the same electrolyte for 30 min. As the electrodes were In clamped ones and from Ag paste with area of 0,3-1 mm$^2$.

### 2.2. Measurements

We investigated the variations of parameters of dynamic current-voltage bipolar (I-V-loops) and unipolar (I-V-curves) characteristics, charge-voltage (Q-V-) loops, and also dynamic transient currents (I-t-curves) induced by continuous and pulse (1-3 sec of duration) relative humidity $H_r$ changes.

Under measuring I-V-loops, I-V-curves and I-t-curves as a load was used the reference resistor and during the measurements of Q-V-loops was used the reference capacitor, as was done under investigations of corresponding characteristics of ferroelectric capacitors [2-4, 7].

The corresponding measurements were performed in the multi-cycle regime under applied a.c. triangular drive voltage $V_d$ and meander voltage $V_0$ in the frequency range 0,1 Hz$\leq$ f $\leq$1 kHz and applied voltage range of $\pm$10 mV $\leq$ V $\leq$ $\pm$10 V.

For minimization of the contribution of polarization reversal currents under I-V-curves examination the unipolar saw-tooth drive voltage $V_d$ was applied.

The temporal changes of investigated characteristics of Ag – PSi – Ag system during and just after wet air pulse and also under restoration of the initial state were also under registration.

## 3. RESULTS AND COMMENTS

Figure 1 and Figure 2 present the current-voltage loops and transient current-time curves before, under and after humidity pulse impact.

The shape of low-voltage I-V-loops at 1 kHz (Figure 1, left) and low-voltage I-t-curves (Figure 2, left) and low $H_r$ value corresponds to equivalent linear sequence-parallel R-C-circuit.

Transformation of I-V-loops and I-t-curves under action of wet air pulse corresponds to decrease of R value and increase of C value under occurrence of apparent voltage R-C-non-linearity. This state with high C and low R values remains during 5-20 sec depending on duration of $H_r$-pulse. Returning to the initial state is accompanied by decrease of degree of R- and C- non-linearity to the initial value.

Under decreasing frequency become apparent R- and C- voltage non-linearity, and at infra-low frequencies the I-V-loops show high R- and C- non-linearity (Fig. 1, right). Under $H_r$ increasing is observed appearance and broadening of hysteresis regions on the positive and negative branches of I-V-loops (Fig. 1, right) with corresponding appearance of characteristic "hump" on I-t-curves (Fig. 2, right). The height of this "hump" is maximal at high $H_r$ and decrease under drying of PSi sample in course of returning to the initial state.

The Q-V-loops obtained under low and high $H_r$ are presented in Figure 3. The increase of $H_r$ leads to the increase of the loop vertical size which corresponds to the increase of the value of electrical charge transferred and to the vertical shift of the loop (see low part of Fig. 3).

Unipolar I-V-curves are presented in Figure 4. Low-voltage I-V-curves under increasing humidity are changed from almost linear to weakly non-linear. Under high $V_d$ the increase of $H_r$ leads to the increase of non-linearity degree and I-V-curves acquire exponential character.

## 4. DISCUSSION

The change of the shape of low-voltage I-V-loops under humidity variation can be characterised by series-parallel R-C-circuit with pronounced R(V) and C(V) dependences.

The main observed peculiarities of I-V-loops can be explained by its simplifying to the parallel R-C-circuit neglecting the effect of series resistor. Since for this R-C-circuit $I(V) = d(CV)/dt + V/R$, under $V = V_0(1 \pm bt)$ with b = const and C = const the current value is $I(V) = \pm V_0(bC + (1 \pm bt)/R)$. So any deviation of I(V) from linearity is connected with existence of some R(V) and C(V) dependences.

The existence of a "hump" on I-V-curves is characteristic for both ferroelectric systems under pulse switching of polarization [7] and semiconductor systems under transfer of injected charge carriers [8]. The unipolar I-V-curves near to exponential ones in the case of injection correspond to a wide spectrum of distribution of centers of capture on energy in the forbidden zone of semiconductor material [8].

The vertical shift of Q-V-loops is connected with the rectification effect in consequence of I-V-loops asymmetry.

The results obtained for the samples PSi are comparable with those obtained for porous metal-oxide ceramics and zeolite-like (of Na-Y type) and silica mezo-porous systems (of MCM-41 type) [2, 3].

It should be pointed out the similarity of view of observed I-V-loops for PSi and I-V- loops for the model semiconductor-ionic $Ag_3AsS_3$ and $Ag_3SbS_3$ [9], for which is characteristic the electrolytic reactions of types of $Ag^+ + e^- = Ag^0$ in under-surface regions.

For the investigated systems the observed peculiarities of I-V loops and I-t curves are connected with adsorption of $H_2O$ vapor, after-dissociation of water molecules, $H_2O \rightarrow H^+ + OH^-$ and ion-electronic transfer in porous conglomerate of PSi and desorption of $H_2O$ as the source of ionic charge carriers. The mechanism of charge transfer in the PSi is connected with the hopping transport of $H^+$ ions by means of switching OH⁻ dangling bonds [10]. Kinetics of these processes determines the shape of corresponding characteristics and peculiarities of their time transformation.

With certain degree of approximation it can be considered that observed times in the consequence "transformation of I-V loops (I-t curves and O-V-loops) – their stabilization in time – restoration" corresponds to characteristic times in the consequence "adsorption-dissociation and transfer – desorption".

## 5. CONCLUSION

For PSi as for an example was shown the efficiency of using the methods of dynamic of electrophysical characterization for studying of changing characteristics of porous materials under fast humidity changes.

The view of high voltage current-voltage curves is characteristic for poling processes in the space charge region similar to that observed in the case of typical ionic conductors.

Observed phases of transformation of investigated electrophysical characteristics reflect the time scale of processes in the consequence "adsorption-dissociation and transfer – desorption".

**References**
1. P. Behrens, "Meso-porous inorganic solids", Adv. Mater., v. 5, No 2, pp. 127-131 (1993).
2. S. L. Bravina, N. V. Morozovsky, G. M. Tel'biz, A. V. Shvets, "Electrophysical characterization of meso-porous structures", Materials of International Conference on Optical Diagnostics Materials and Devices for Opto-, Micro and Quantum Electronics, SPIE, Kiev, (1999).
3. S. L. Bravina, N. V. Morozovsky, E. G. Khaikina, "Temperature and humidity sensors with response to frequency conversion based on porous ceramics", Materials of 12th IEEE International Symposium on the Application of Ferroelectrics (ISAF, Hawaii, 2000).
4. S. L. Bravina, N. V. Morozovsky, "Porous Si thin layer based humidity sensitive elements". Materials of Second Open French-Ukrainian Meeting on Ferroelectricity and Ferroelectric Thin Films (Dinard, 2002).
5. S. L. Bravina, I. V. Blonsky, N. V. Morozovsky, V. O. Salnikov, "Thermal characterization of porous Si as IR-sensor substrate material", Ferroelectrics. v. 254, pp. 65-76 (2001).
6. A. Foucaran, B. Sorli, M. Garcia, F. Pascal-Delannoy, A. Giani, A. Boyer, "Porous silicon layer coupled with thermoelectric cooler: a humidity sensor", Sens. Actuators, A 79, pp. 189-193 (2000).
7. *J. C. Burfoot, Ferroelectrics. An introduction to the Physical Principles, Princeton-New Jersey-Toronto, 1967.*
8. M. A. Lampert and P. Mark, *Current Injection in Solids,* Academic, New-York, 1970.
9. S. L. Bravina, N. V. Morozovsky, "Switching effect in the systems metal-$Ag_3AsS_3(Ag_3SbS_3)$-metal", Phys. and Techn. Semicond., v. 17, No 5, pp. 824-828 (1983).
10. D. Stievenard, D. Deresmes, "Are electrical properties of an aluminium porous silicon junction governed by dangling bonds?", Appl. Phys. Lett., v. 67, No 7, pp.1570-1572 (1995).

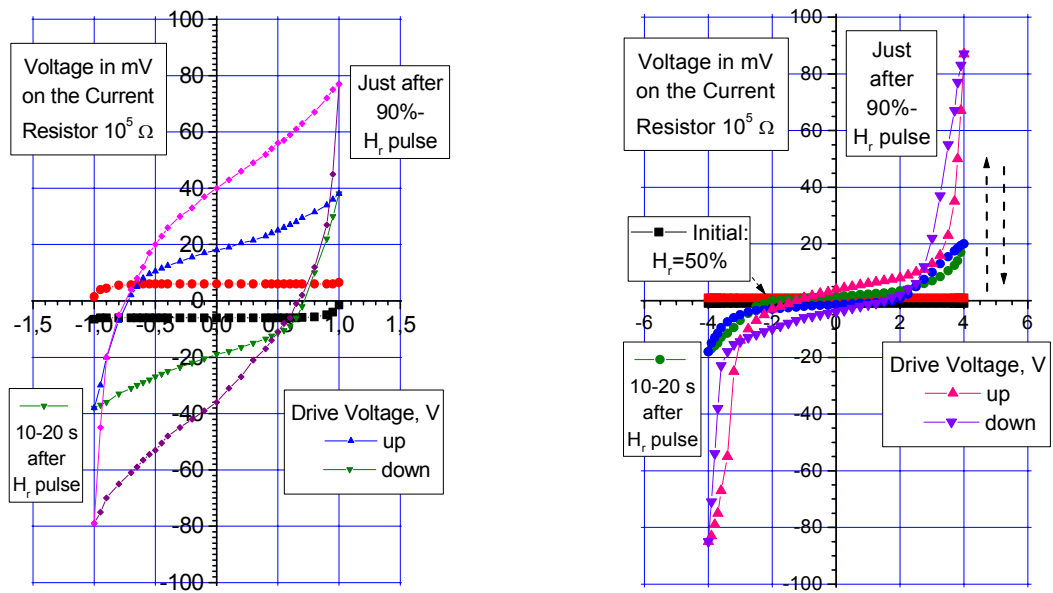

FIGURE 1. Current-voltage loops before, under and after humidity pulse impact at different amplitudes and frequencies of drive voltage (left: 1 V, 1 kHz and right: 4 V, 1 Hz).

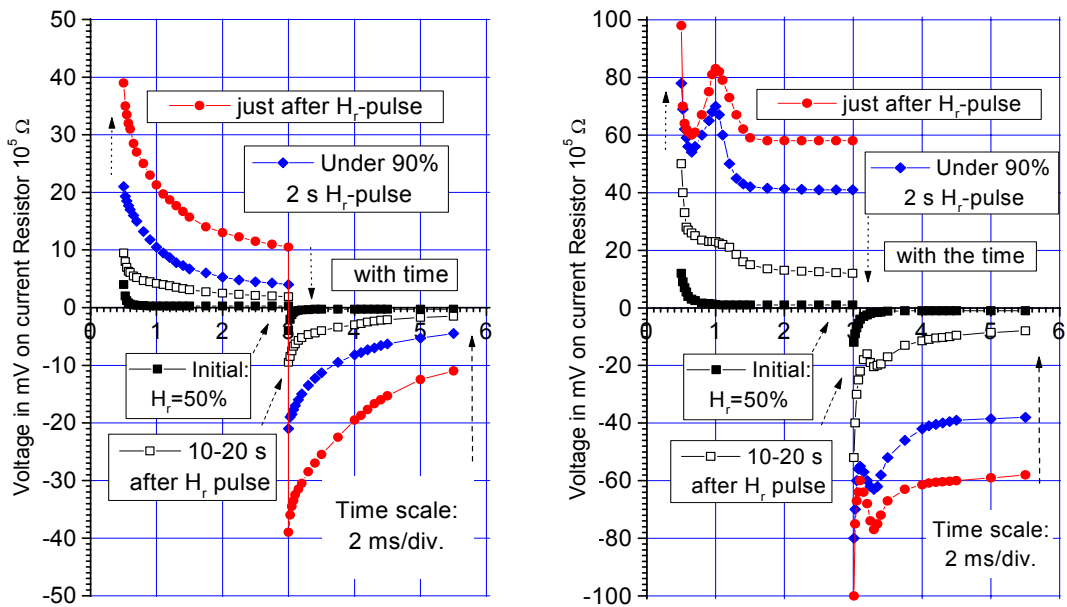

FIGURE 2. Transient current-time curves before, under and after humidity pulse impact at different amplitudes of 10 Hz meander voltage (left: 1 V and right: 4 V).

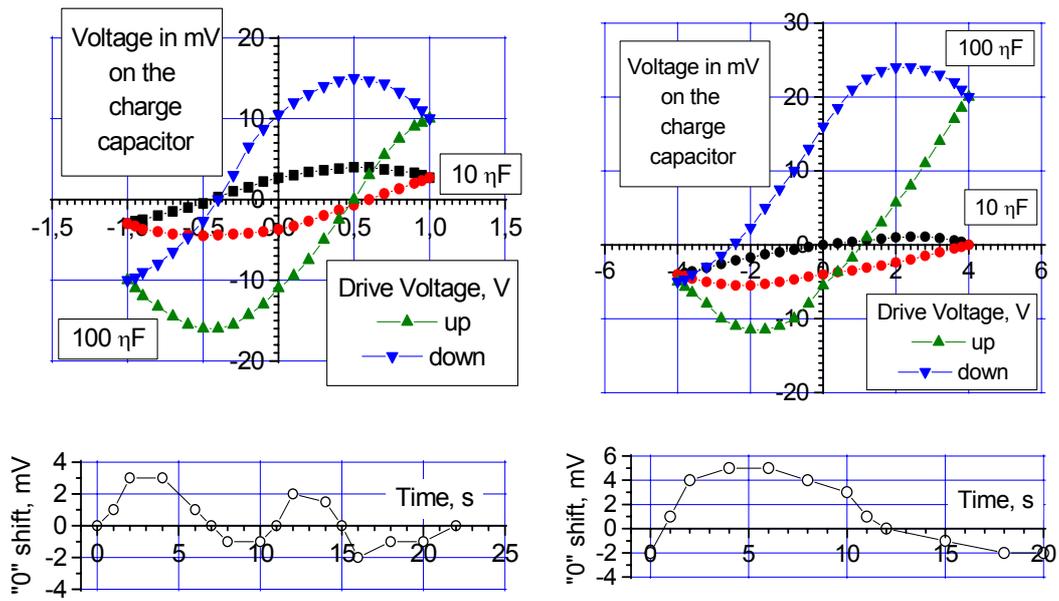

FIGURE 3. Charge-voltage loops obtained under low and high humidity at different amplitudes of 10 Hz drive voltage (left: 1 V and right: 4 V) (upper part) and changes of the position of the centre of loop after humidity pulse impact (lower part).

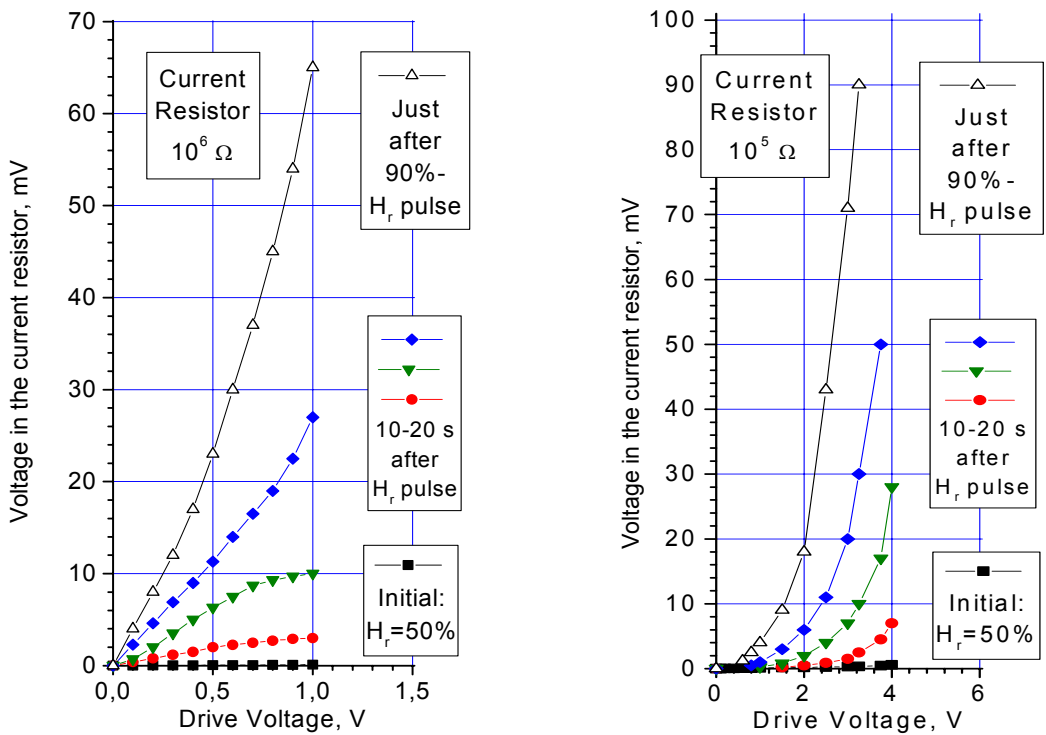

FIGURE 4. Unipolar current-voltage curves before, just after and after humidity pulse impact at different amplitudes of 1 Hz drive voltage (left: 1 V and right: 4 V).